\documentclass[11pt]{article}
\pdfoutput=1
\usepackage{epsfig}
\usepackage{graphicx}
\usepackage{multicol}
\usepackage{amsmath,amssymb}
\usepackage{hyperref}
\numberwithin{equation}{section}
\usepackage{cite}
\usepackage{graphicx,wrapfig,color}
\usepackage{multicol}
\usepackage{tabularx}

\newcommand{\bea}{\begin{eqnarray}}
\newcommand{\eea}{\end{eqnarray}}
\newcommand{\be}{\begin{equation}}
\def\bel#1{\begin{equation} \label{#1}}
\newcommand{\ee}{\end{equation}}
\newcommand{\ba}{\begin{align}}
\newcommand{\ea}{\end{align}}
\newcommand{\comments}[1]{}

\def\pref#1{(\ref{#1})}

\def\cv{{\mathcal{V}}}
\def\hp{{\hat{\phi}}}
\begin{document}
\begin{titlepage}
\comments{ \hfill{preprint number}}
\vskip 1 cm
\begin{center}
{\Large \bf 
Fibre inflation and precision CMB data
}
\vskip 1.5cm

\begin{center}
Sukannya Bhattacharya$^{a}$,  Koushik Dutta$^{b,c}$,  Mayukh Raj Gangopadhyay$^{d}$ ,  \\
Anshuman Maharana$^{e}$, Kajal Singh$^{e}$
\end{center}

\vskip 0.9 cm
{\textsl{
$^{a}$Theoretical Physics Division, Physical Research Laboratory, Navrangpura,\\ Ahmedabad 380009, India \\
$^{b}$ Indian Institute of Science Education And Research Kolkata,\\
Mohanpur, West Bengal 741 246, India \\
$^{c}$Theory Divison, Saha Institute of Nuclear Physics,
HBNI,1/AF Bidhannagar,\\ Kolkata 700064, India \\
$^{d}$Centre for Theoretical Physics, Jamia Millia Islamia,\\ New Delhi 110025, India\\
$^{e}$Harish-Chandra Research Institute, HBNI, Jhunsi,\\ Allahabad, Uttar Pradesh 211019, India \\}
}
\end{center}

\vskip 0.6cm

\begin{abstract}

       Generic features of models of inflation obtained from string compactifications are the correlations between the model parameters
and the postinflationary evolution of the universe. Thus, the postinflationary evolution depends on the inflationary model parameters and accurate inflationary predictions require that this be incorporated in the evolution of the primordial spectrum.  The fibre inflation model is a promising model
of inflation constructed in type IIB string theory. This model  has two interesting features in its postinflationary evolution. The reheating temperature of the model is directly correlated with the model parameters. The model also necessarily predicts some dark radiation, which can be sizable for certain choices of discrete parameters in the model. We analyze this model in detail using publicly available codes - {\sc{ModeChord}} and {\sc{CosmoMC}} with the latest \textit{Planck+BICEP2/Keck array} data to constrain the model parameters and $N_{\rm pivot}$ (the number of $e$-foldings between horizon exit of the CMB pivot mode and the end of inflation).  We also carry out the
same analysis using the publicly available code {\sc{Cobaya}}. We find the results of both the analysis to be in agreement.
Our analysis provides the basic methods necessary to extract precise inflationary prediction in string models incorporating correlations between model parameters and postinflationary evolution.

\end{abstract}

\vspace{1cm}

\let\thefootnote\relax\footnotetext{email: {$\mathtt{sukannya@prl.res.in, \ koushik@iiserkol.ac.in, \ mayukh@ctp-jamia.res.in,}$}}
\let\thefootnote\relax\footnotetext{\phantom{email:} {$\mathtt{anshumanmaharana@hri.res.in, \ kajalsingh@hri.res.in}$}}

\end{titlepage}
\pagestyle{plain}
\setcounter{page}{1}
\newcounter{bean}
\baselineskip18pt
%



%
%
\section{Introduction}

The inflationary paradigm provides an extremely  attractive explanation  for the observed spectrum and inhomogeneities in the cosmic microwave background (CMB)~\cite{p18inf}.
Since observations in the future are likely to minutely probe the CMB~\cite{Aba}, it is important to develop a systematic understanding of the methodology for extracting
highly accurate predictions for inflationary models. The simplest method is to  parametrize  primordial perturbations with a set of empirical variables
such as $A_{s}$ (the strength of the power spectrum), $n_s$ (the scalar tilt), $r$ (the tensor-to-scalar ratio), $f_{\rm NL}$ (parametrizing the non-Gaussianity) etc. The best-fit
values of these are obtained by evolving the primordial fluctuations and comparing with observations of the CMB. Given a model of inflation, one can also compute
the  functional form of the primordial fluctuations in terms of the parameters of the model. One then requires that the predictions for the empirical 
parameters are in the best-fit regions, determined by the evolution of the initial perturbations. Note that this is intrinsically a two-step process where  the empirical
parameters characterizing the primordial perturbations act as the matching points between observation and theory.

   On the other hand, if one wants to confront a particular model of inflation with data, a more comprehensive method is to treat the model parameters as inputs
for the cosmological evolution and directly determine the best-fit regions for these parameters \cite{Jm1, modecode} (see also \cite{M1}). This approach is particularly well suited if one is considering
models which arise from a fundamental theory (such as string theory). In this case,  one naturally expects various correlations between the model parameters
and the postinflationary evolution of the universe. Thus, the postinflationary evolution depends on the model parameters and accurate inflationary predictions
require that this be incorporated in the evolution of the primordial spectrum.  Inflationary predictions of any model are sensitive to higher derivative corrections in the effective action. Hence, theories of quantum gravity are
 the appropriate setting to carry out inflationary model building. In this light, on can expect to use precision cosmology to confront models of quantum gravity with observations. Work in this direction, to explicitly constrain model parameters incorporating the correlations between model parameters and
 the postinflationary history was initiated in \cite{kcmb} in the context of the K\"ahler moduli inflation model \cite{fq}.

      Fibre inflation \cite{f1} is a promising model of inflation set in IIB string theory. Phenomenologically, the model is interesting as it predicts a value of the tensor to
scalar ratio ($r > 0.005$) which can be observationally verified with experiments planned in the near future. Thus, it is timely to carry out a detailed study of the model predictions.
 As we will discuss in detail in the next section,  the model has two interesting
features in its postinflationary evolution. The reheating temperature of the model is directly correlated with the model parameters. The model also necessarily
predicts some dark radiation, which can be sizable for certain choices of discrete parameters in the model. In this paper, we use {\sc{Modechord}}~\cite{modecode} and {\sc{CosmoMC}}~\cite{mc} (and also give an independent analysis using {\sc{Cobaya}} \cite{cobaya}), incorporate these
features in the postinflationary evolution and thereby perform a detailed analysis of the model predictions.\footnote{
The two independent analyses give results which are in agreement} The basic philosophy of constraining the
model parameter space using precision cosmology is the same as that in \cite{kcmb}.

 These two features are also expected to be generic
in string constructions.\footnote{Another generic feature is the epochs in the postinflationary history in which the energy density is dominated
by cold moduli particles. Its effect on inflationary predictions has been studied in detail in \cite{gkkane}.}  For a discussion of dark radiation in string models, see e.g. \cite{d1}. Thus, our analysis can serve as a template for the analysis of most string models. The effects of presence of dark radiation on cosmological observations are studied in \cite{darkexp}.

   Recently, the predictions of fibre inflation and their relationship to post inflationary dynamics have been analyzed in\footnote{For a complimentary approach
 see \cite{cab}.} \cite{fr}. Our work develops this analysis, 
systematically incorporating the relationship between the model parameters and the post inflationary dynamics making use of the above-mentioned publicly 
available packages. This allows us
to obtain a detailed understanding of the model predictions. 

  This paper is structured as follows. In Sec.~\ref{srev}, we review some basic aspects of fibre inflation; in Sec.~\ref{sm} we discuss our methodology and perform
our analysis; in Sec.~\ref{scon} we discuss our results and conclude.

\section{Review of Fibre Inflation}
\label{srev}

  The fibre inflation model is set in the large volume scenario \cite{LVS} for moduli stabilization of IIB flux compactifications. Here, we briefly review aspects of the model that will be needed for our analysis and refer the reader to \cite{f1, fr,  f2} for further details.\footnote{We follow the conventions and
  notation of \cite{fr}.} The relevant dynamics during the inflationary epoch is that of the K\"ahler moduli fields{\footnote{The complex structure moduli are fixed by fluxes \cite{GKP}.}  of the Calabi-Yau manifold associated with the compactification.  
The K\"ahler moduli are flat at tree level, but acquire a potential as a result of nonperturbative corrections to the superpotential, loop, and $\alpha'$ corrections to
the K\"ahler potential. The construction of fibre inflation models involves Calabi-Yau manifolds with at least three K\"ahler moduli{\footnote{We will denote the K\"ahler moduli
as $T_{i} = \tau_{i} + \theta_{i}$, with $\tau_{i}$ being a geometric modulus and $\theta_{i}$ its axionic partner.}}:
\begin{itemize}
\item $T_{1} = \tau_1 + i \theta_1$. For this field, the geometric modulus $\tau_1$ corresponds to the volume of a $T^{4}$ or $K3$ fibred over a $\mathbb{P}^1$ base. The field
$\tau_1$ plays the role of the inflaton in the model.
\item $T_{2} = \tau_2 + i \theta_2$. Here the geometric modulus corresponds to the volume of the  base.
\item $T_{3} = \tau_3 + i \theta_3$. Here, the geometric modulus corresponds to the blow-up of a pointlike singularity. Nonperturbative effects on this cycle play an 
important role in moduli stabilization.
\end{itemize}
 The volume of the compactification can be expressed in terms of the volumes of the geometric moduli as
 \bel{vfor}
   \cv =  \alpha \left(  \sqrt{\tau_1} \tau_2  - \gamma \tau_3^{3/2} \right),
 \ee 
where $\alpha$ and $\gamma$ are order one constants determined by the intersection numbers of the four cycles.

 The potential developed as a result of the effects described above can be expanded in an inverse volume expansion. At   order ${\mathcal{V}}^{-3}$,  the geometric moduli
 $\tau_2$ and $\tau_3$ and the axion $\theta_3$ are stabilized. Loop effects at  order  ${\mathcal{V}}^{-10/3}$ provide a potential for the field $\tau_1$. This takes the
 form (in Planck units)
 \begin{equation}
 V(\tau_1)  = \left( g_s^2 { A \over \tau_1^2 }  -  { B \over  \sqrt{\tau_1} } + g_s^2 { C \tau_1 \over \cv^2} \right) { W_{0}^2 \over \cv^2},
 \end{equation}
 %
 where $W_0$ is the vacuum expectation value of the Gukov-Vafa-Witten superpotential and
 $$
   A =  \left( c_1^{\rm KK} \right)^2 \phantom{abcdef}  B = 2\alpha c^{\rm W} \phantom{abcdef}  \phantom{abcd} C =  2 (\alpha c_2^{\rm KK})^2
 $$
 with $c_1^{\rm KK}$, $c_2^{\rm KK}$, and $c^{\rm W}$ depending on the underlying compactification and fluxes. After incorporation of effects so that the minimum is a Minkowski one, canonical normalization of $\tau_1$,
and  shifting the zero of the field to its minimum, the potential for the canonically normalized inflaton  field $(\hat{\phi})$ is
 \bel{ipot}
   V = V_{0} \left( 3 - 4 e^{- k \hat{\phi} } + e^{-4 k \hat{\phi}}  + R \left( e^{2 k \hat{\phi}}  - 1 \right) \right),
\ee
 where
\bel{pdef}
 k = {1 \over \sqrt{3} }, \phantom{abcd} V_{0} =   { g_s^{1/3} W_0^2 A  \over 4 \pi \lambda^2 } \phantom{abc} \textrm{with} \phantom{abc} \lambda =  \left({4A  \over B} \right)^{2/3}, \phantom{abcd} \textrm{and} \phantom{abcd}
  R =  16 g_{s}^4 { AC \over B^2}
\ee
 The inflationary trajectory is such that $\hp$ rolls from positive values towards its minimum at zero. Note that $R \propto g_{s}^4$ and hence is naturally small. The potential
 has two inflection points:  $\hp_{\rm ip}^{(1)} \sim k \ln 4$ and $\hp_{\rm ip}^{(2)} \sim - k \ln R$. The second inflection point occurs as a result of competition between
 the positive exponential and the negative ones. Inflation occurs when the field lies between the two inflection points. If the value of $R$ is small, $R <  2 \times 10^{-6}$,
 then the horizon exit of the CMB modes takes place at a field value $(\hp_{*})$ which is much less than the second inflection point $\hp_{*} \ll \hp^{(2)}_{\rm ip}$, and the positive exponential term can be neglected. In this regime, a robust prediction of the model is a relationship between the spectral tilt $(n_s)$ and the tensor-to-scalar ratio $(r)$
 \bel{rob}
    r = 6 (n_s - 1)^2
\ee
 On the other hand, for higher values of $R$, the horizon exit of CMB modes takes place at a point which is closer to the second inflection point; the positive
 exponential term has to be incorporated in the analysis. With an increase in the value of $R$, the model predicts higher values of $n_s$ and $r$. Also, the relationship
 \pref{rob} is broken.
 
    The reheating epoch in fibre inflation models has been examined in detail in \cite{fr}. After the end of inflation, the inflaton oscillates about its minimum and decays
perturbatively, which is supported by the full numerical analysis of the evolution of the scalar field after inflation \cite{Ant} (a semianalytic approach \cite{Gu}
has suggested the possibility of a  preheating epoch, but the evidence from the full numerical study is that the process is perturbative) . The dominant decay channels are visible sector gauge bosons, visible sector Higgs, and ultralight bulk hidden axionic fields (which act as dark radiation). The total visible sector and the hidden sector decay widths are given by

\bel{wii}
  \Gamma^{\rm vis}_{\rm \hp}  =  12 \gamma^{2} \Gamma_0 \phantom{abc} \textrm{and} \phantom{abc}  \Gamma^{\rm hid}_{\rm \hp} = { 5 \over 2 } \Gamma_0~,
\ee
 where $\Gamma_0 = { 1 \over 48 \pi} {m_{\hp}^3 \over M_{\rm pl}^2 }$, and 
 \bel{gdef}
    \gamma = 1 + \alpha_{\rm vis} { h(F_1) \over g_s},
 \ee
  where $\alpha_{\rm vis}$ is the high-scale visible sector gauge coupling $(\alpha_{\rm vis}^{-1} \sim 25)$, and $h(F_1)$ depends on the $U(1)$ flux threading of the D7 brane
  on which matter fields are localized. It vanishes for zero flux, and is an order one quantity as the flux quanta is increased.\footnote{More precisely,
  $ h(F_1) = {1 \over 2} k_{112} n_2^2$, where $k_{112}$ is a triple intersection number involving the two-cycles dual to the three four-cycles of the Calabi-Yau and
  $n_i$ the integral coefficients of the expansion of the gauge flux in terms of these dual cycles \cite{jl}.}
  
    Given the widths in \pref{wii} the prediction for dark radiation is easily computed. One finds 
$$
 \Delta N_{\rm eff} = {0.6 \over \gamma^2 }~.
$$
Thus the model necessarily predicts some dark radiation. The prediction is high in the absence of any gauge flux, and can be sizable for small values
of the flux quanta. Recall, that  the analysis of Planck prefers higher values of $n_s$ in the presence of dark radiation. As we have discussed earlier,
this can be obtained with higher values of the parameter $R$ in the inflationary potential \pref{ipot}.

  Finally, we come to the number of $e$-foldings before horizon exit.  This is given by (see  e.g. \cite{ll, planck2015})
\bel{nef}
  N_{\rm pivot} = 57 + {1 \over 4} \ln r +   {1 \over 4}\ln \left( \rho_{*} \over \rho_{\rm end} \right) + 
  { {1 - 3 w_{\rm rh} } \over 12 (1 + 3 w_{\rm rh}) } \ln \left( {\pi^{2} \over 45 }g_{*} (T_{\rm rh}) \right) - 
  {1 \over 3} { {1 - 3 w_{\rm rh} } \over  (1 + 3 w_{\rm rh}) } \ln \left( M_{\rm inf} \over T_{\rm rh} \right),
   \ee
 where $\rho_{*}$ and $\rho_{\rm end}$ are the energy densities of the universe at the time of horizon exit of the pivot scale $k_*^{-1}$ and at the end of inflation. Here, $w_{\rm rh}$
 is the average equation of state during the reheating epoch, $g_{*} (T_{\rm rh})$ is the number of relativistic degrees of freedom at the end of reheating and
 $T_{\rm rh}$ is the reheating temperature.  
   The reheating temperature can be obtained from Eq.~\pref{wii} as 
 \bel{trh}
   T_{\rm rh} = 0.12 \gamma m_{\hp} \sqrt{ m_{\hp} \over M_{\rm pl} }, 
 \ee
where $m_{\hp}$ is the mass of $\hp$ about the minimum at $\hp=0$ in \pref{ipot}. Note that this implies that the number of $e$-foldings before horizon exit in the
model is correlated with parameters in the potential and the amount of dark radiation (although the dependence on the amount of dark radiation is very mild
as $\gamma$ is an order one quantity). Since the inflaton decays perturbatively and has a long lifetime, we take $w_{\rm rh} = 0$.

Before closing this section, we would like to emphasize that, as in many string models, in fibre inflation there is a direct correlation between $N_{\rm pivot}$ and
 the parameters in the inflationary potential. The model has the interesting feature that for certain discrete choices in the parameter space,
 a considerable amount of dark radiation is predicted. Furthermore, the model's prediction for the tensor-to-scalar ratio $r$ is in the right ballpark to be probed by upcoming CMB B-mode
 observations. Given this, a detailed analysis of the model which takes into account the above considerations is very well motivated, and it is the primary
 goal of this paper.
 
\section{Methodology and Results}
\label{sm}

In this section, we discuss our methodology for parameter estimation and report our results. First, we note that the potential in Eq.~\eqref{ipot} has two parameters $V_0$ and $R$, which themselves depend on some fundamental parameters (such as the volume of the compactification and $W_0$). Thus, these two parameters broadly control the inflationary perturbations. But, these two parameters also control the postinflationary history via Eq.~\eqref{trh}. On the other hand, the parameter $\gamma$ controls the amount of dark radiation $\Delta N_{\rm eff}$.

 \begin{figure}[h]
\centering
\includegraphics[width=0.4\textwidth]{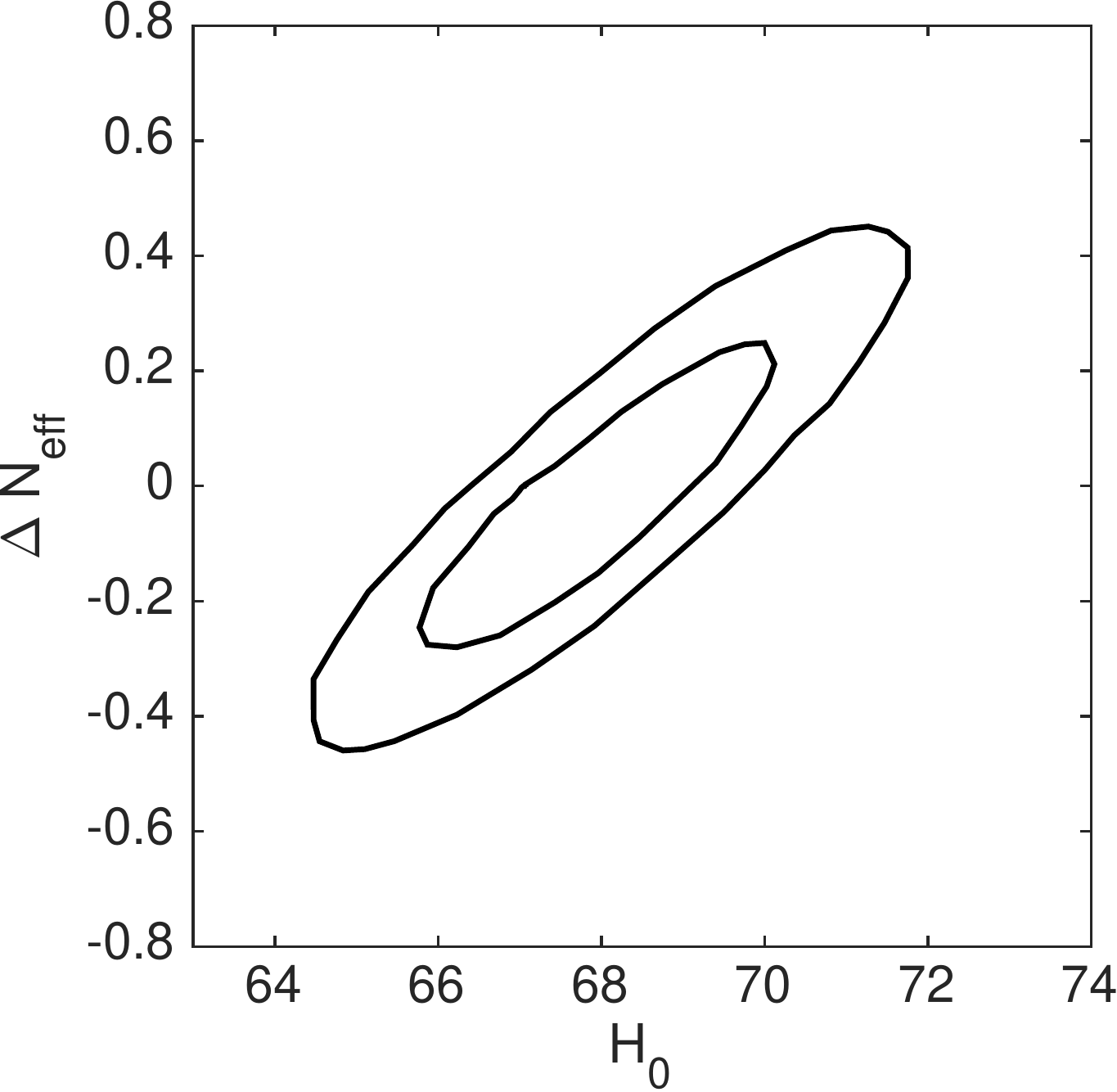}
\caption{Marginalized posterior distributions in the $\Delta N_{\rm eff}$ and the $H_0$ plane [the contours correspond to the $1\sigma$ and $2\sigma$ confidence limits (C.L.)]. Here, $H_0$ is plotted in 
units of $\text{km} \ \text{s}^{-1} \text{Mpc}^{-1}$. }
\label{fig:f1}
\end{figure}}
 \begin{figure}[h]
\centering
\includegraphics[width=0.4\textwidth]{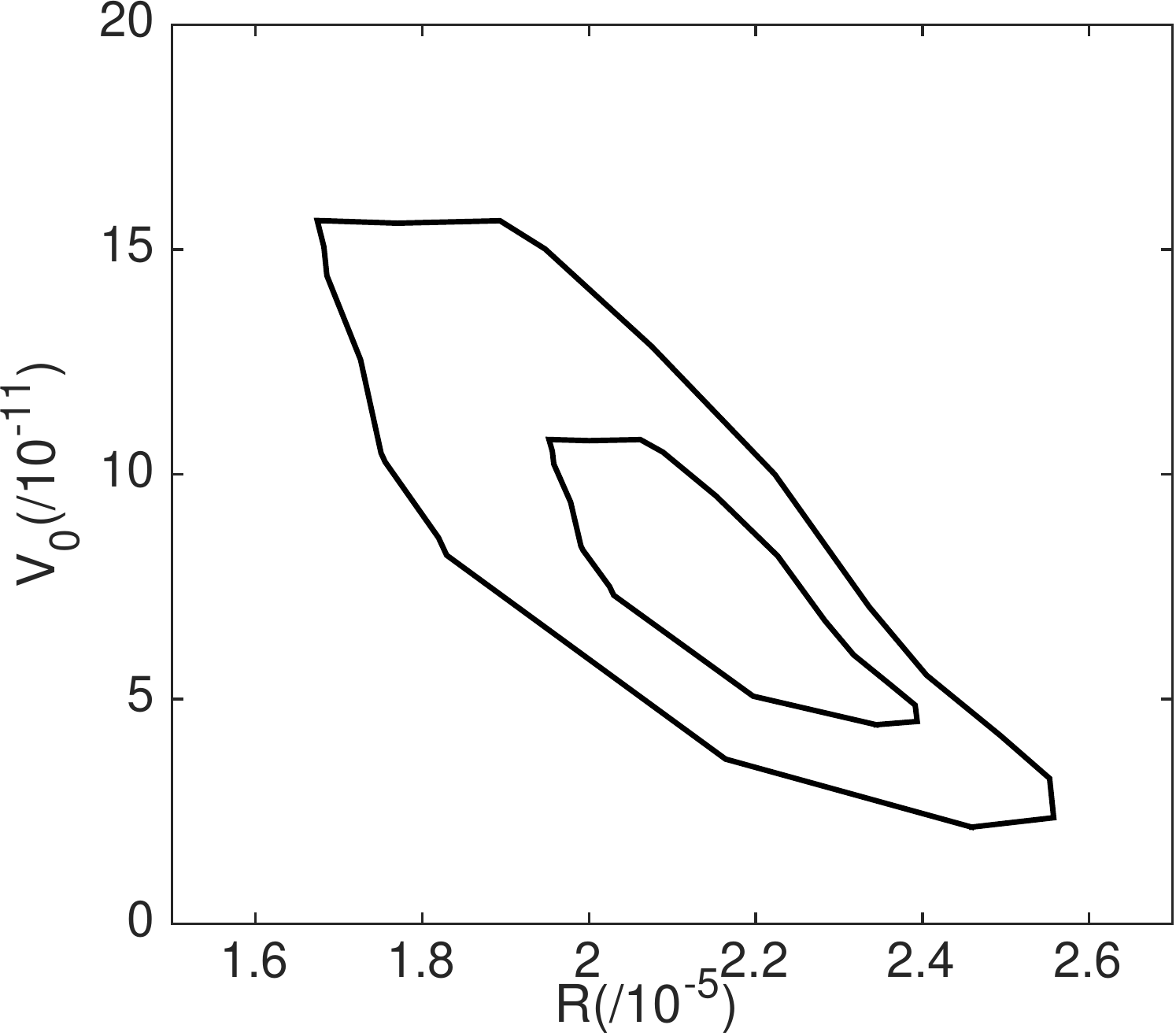}
\caption{Favored region of the model parameter $R$ with respect to the scale of the inflation $V_0$ (in reduced Planck units). }
\label{fig:f5}
\end{figure}
%

For given values of the model parameters $R$ and $V_0$, and $N_{\rm pivot}$, we have evaluated the cosmological perturbations  by using {\sc{ModeChord}}  \cite{modecode} (together with {\sc CosmoMC} through {\sc{Multinest}}~\cite{multinest}) without assuming slow-roll conditions. Along with these parameters, we have also varied $\gamma$ which controls the amount of dark radiation produced. As usual, the Boltzmann solver CAMB \cite{cambb} is used to evaluate the two-point correlation functions for temperature and polarization, and then the model parameters are estimated and the goodness of fit is determined using {\sc{CosmoMC}}~\cite{mc}. The likelihoods used here are from \textit{Planck 2018 TT+TE+EE + lowP + lensing} and \emph{Planck+BICEP2/Keck array} joint analysis~\cite{keck}. The model parameters are then inferred from the chains of the simulation using the code in \cite{matlab}.

In Fig.~\ref{fig:f1}, the dark radiation allowed by \textit{Planck'18} data is plotted with respect to  the Hubble constant (the contours correspond to the 1$\sigma$ and 2$\sigma$ C.L.). Note that, $\Delta N_{\rm eff}$ represents the extra presence of radiation with respect to the  theoretically expected $N_{\rm eff}\sim 3.046$ from the Standard Model (SM) of particle physics. We see that $\Delta N_{\rm eff} = 0$ is fully consistent with the data. In Fig.~\ref{fig:f5},  model parameters $R$ and $V_0$ are plotted against each other. We note that similar constraints of  the model parameter space in the
context of K\"ahler moduli inflation were obtained in \cite{kcmb}.

\begin{figure}[h]
\centering
\includegraphics[width=0.5\textwidth]{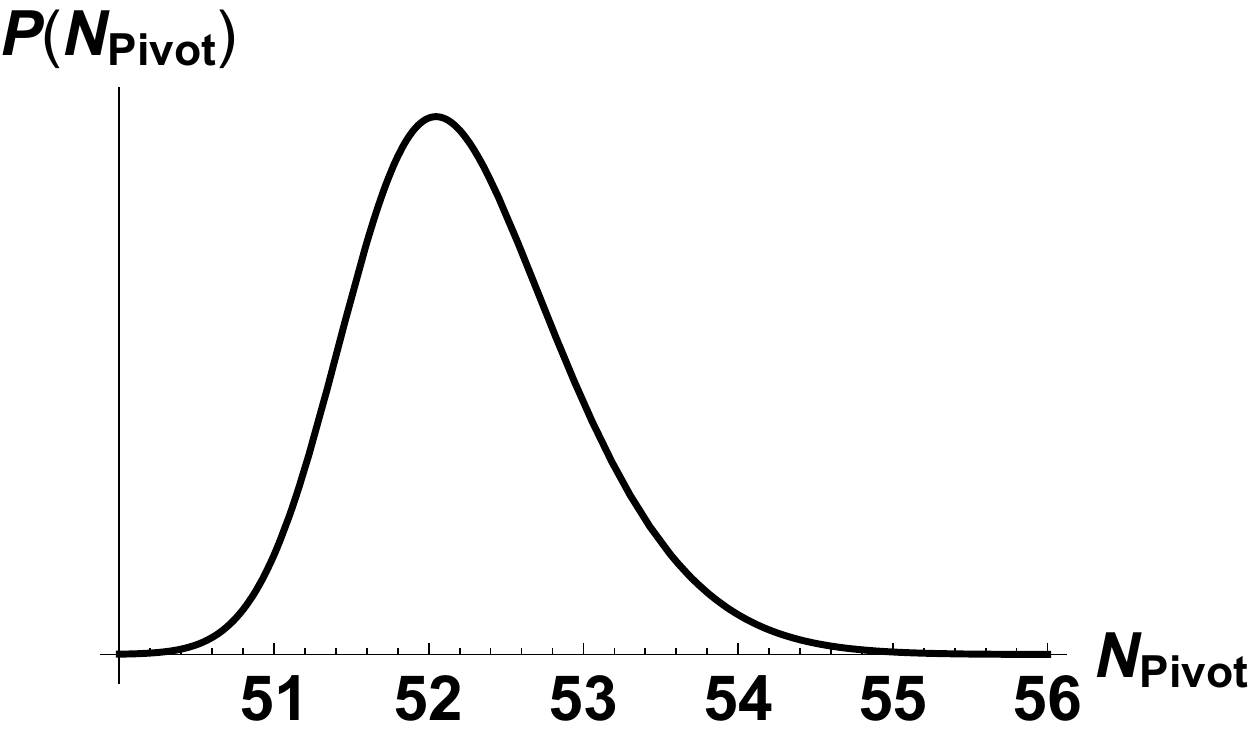}
\caption{1D probability distribution of the number of $e$-foldings $N_{\rm pivot}$.}
\label{fig:f3}
\end{figure}

\begin{figure}[h]
\centering
\includegraphics[width=0.5\textwidth]{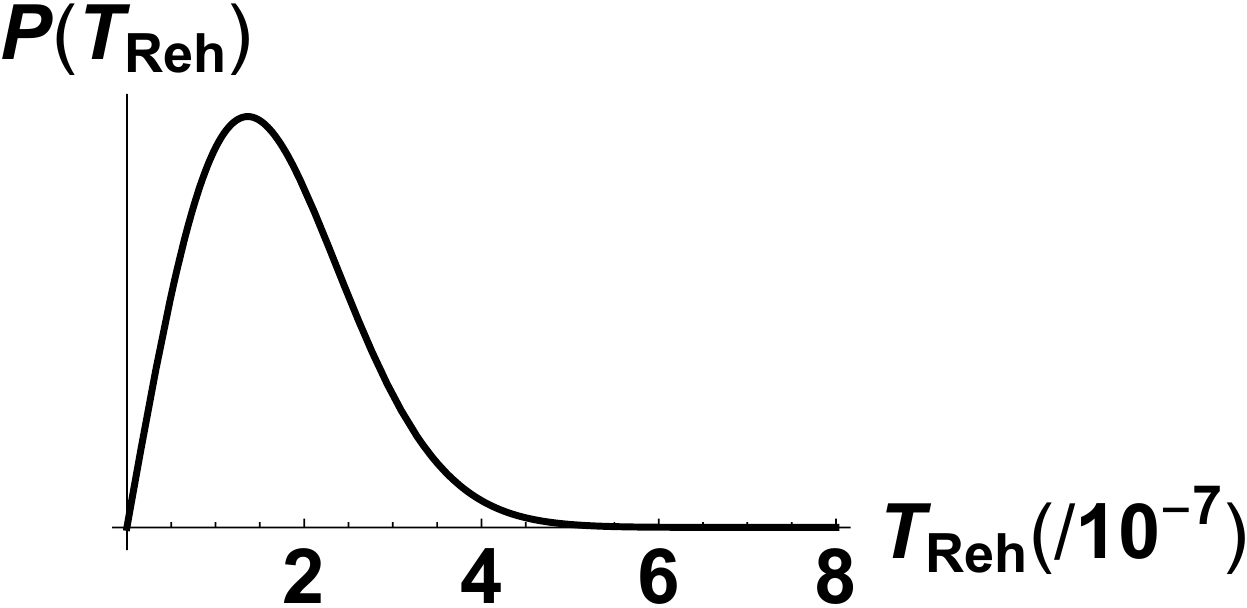}
\caption{1D probability distribution of the reheating temperature $T_{\rm rh}$ (in reduced Planck Units).}
\label{fig:f4}
\end{figure}

 In Fig.~\ref{fig:f3}, the posterior probability distribution of the number of $e$-foldings is plotted, and the central value is found to be around $53$, which is quite close to the estimate in \cite{fr}. Finally, in Fig.~\ref{fig:f4}, we have plotted the posterior probability distribution for the reheating temperature, and the most probable value is around $10^{11}$ GeV.  A summary of the results for the main simulation is given in Table~\ref{T1}. Interestingly, the central value of $n_s$ here has a small shift, $n_s\simeq 0.9691$, as compared to the one obtained from the Planck analysis $n_s= 0.9649$.

 A rough check of our results can be done as follows.  One can take the central values of the $R,V_0$ 
 in Table~\ref{T1} and consider the model potential \pref{ipot} with these values for the parameters. 
 Taking derivatives of this potential at the point in field space corresponding to the central value of $N_{\rm pivot}$ in Table~\ref{T1},
 one finds $n^{\rm cen}_s=  0.986$ and $r^{\rm cen}= 0.0092$  (where the superscript ``cen'' indicates that these quantities 
 are computed from a potential function constructed with the central values obtained from our analysis). We note that these
 central values are close to $n_s$ and $r$ obtained from our simulations (which involves a full statistical sampling over the model parameters): $r^{\rm cen}$ is very close to the value of $r$ in Table~\ref{T1},
 while $n_s^{\rm cen}$ and $n_s$ agree approximately at the $1\sigma$ level.  We take these agreements as consistency checks of 
 our numerics.

\comments{
The central values of the model parameter $R$ and $N_{\rm pivot}$  from Table~\ref{T1} along with the relations in Eq.s~\eqref{ns_theory} and~\eqref{r_theory} give $n_s=0.986$ and $r=0.0092$. Comparing with the values of $n_s$ and $r$ in Table~\ref{T1}, it is evident that the values of $r$ match very well ($<1-\sigma$ consistent), whereas the values of $n_s$ are consistent at just over $1-\sigma$. However, if the $1-\sigma$ allowed range for $R$ and $N_{\rm pivot}$ are inserted in Eq.s~\eqref{ns_theory} and~\eqref{r_theory}, the resulting values of $n_s$ and $r$ become $1-\sigma$ consistent with those quoted in Table~\ref{T1}. This points towards the reliability of the correlation between $\Delta N_{\rm eff}$ and $\gamma$, using which the full numerical simulations were executed.}

  Note that our results point to a very small amount of dark radiation. To compare with
the case with sizable dark radiation, a run for a fixed value of $\Delta N_{\rm eff}=0.6$, corresponding to the theoretically well-motivated case of $\gamma=1$, was carried out (here, $N_{\rm pivot}$ was sampled as a function of the model parameters as given by \pref{nef}, 
with the reheating temperature taken to be as in \pref{trh} with $\gamma =1$). The best-fit $\Delta \chi ^2$ for the results in Table~\ref{T1} and for the $\gamma=1$ case are, respectively, 2 and 18. This is expected because Table~\ref{T1} points to the best-fit value $\Delta N_{\rm eff}=0.00041$, which corresponds to large $\gamma$.  Overall, a small $\Delta N_{\rm eff}$ is preferred.

To check if there is any systematic error due to the simulation procedure adopted up to now,  we have investigated fibre infation using another independently developed publicly available Markov Chain Monte Carlo (MCMC) tool named {\sc{Cobaya}}\cite{cobaya}. {\sc{Cobaya}} implements the BOBYQA algorithm \cite{bobyqa}. It is interfaced with the {\sc{Polychord}} nested sampler; thus, coupling {\sc{ModeChord}} with the {\sc{Cobaya}} becomes much easier to implement. Along with the standard Monte Carlo samples, {\sc{Cobaya}} uses an importance-reweighting method, which makes the computation much faster and more efficient. In the end, both MCMC simulation techniques give self-consistent results, but we found it to be a good addition in the analysis, keeping in mind the sensitivity of the results quoted here. We conclude that the end results from the two independent simulations are almost independent of the MCMC sampler adopted. This indicates the robustness of our results. The results obtained using {\sc{Cobaya}} are very consistent with the results quoted in Table~\ref{T1}. For the sake of completeness, we have quoted the results obtained using {\sc{Cobaya}} in Table~\ref{T4}.

\begin{table}[hbt!]
\centering
\begin{tabular}{|m{2cm}  | m{3cm}|   m{2.0cm}|}
\hline
Parameters    &  Central value   & $1\sigma$   \\
\hline
    $R(/10^{-5})$         & 2.1451           &$^{+0.0979}_{-0.0678}$ \\
\hline

 $V_0(/10^{-11})$         & 5.66           &$^{+4.51}_{-1.01}$  \\
\hline  

   $\Delta N_{\rm eff}$     &  0.00041          &$^{+0.21}_{-0.20}$ \\
\hline
$A_s(/10^{-9})$       & 1.300          & $^{+0.950}_{-0.350}$   \\
\hline
 $H_0$          & 68.01               &$^{+1.81}_{-3.27}$\\
\hline  
       
  $n_s$          &  0.9691                       & $^{+0.0128}_{-0.0108}$ \\
\hline           
     
$r $        &0.0093            &$^{+0.0005}_{-0.0006}$\\
           
\hline
            
$N_{\rm pivot}$             &53.26            &$^{+1.58}_{-0.51}$\\
 \hline 
 $T_{\rm rh}(/10^{-7})$             &1.91            &$^{+0.74}_{-0.08}$ \\
 \hline 
 
\end{tabular}
\caption{
Constraints on the model parameters and the cosmological parameters. We used the data combination \textit{Planck'18~TT+TE+EE+ low P +lensing + BKPlanck15.} All dimensionful quantities are in reduced Planck  units.}
\label{T1}
\end{table}

\comments{
\begin{table}[hbt!]
\centering
\begin{tabular}{|m{2cm}  | m{3cm}|   m{2.0cm}|}
\hline
Parameters    &  Central value   & $1\sigma$   \\
\hline
    $R(/10^{-5})$         & 1.98           &$^{+0.0979}_{-0.0678}$ \\
\hline

 $V_0(/10^{-11})$         &6.12         &$^{+4.561}_{-1.111}$  \\
\hline
$A_s(/10^{-9})$       & 1.275         & $^{+0.999}_{-0.211}$   \\
\hline
 $H_0$          & 67.52              &$^{+1.01}_{-1.12}$\\ 
\hline  
     $n_s$          &  0.9680                       & $^{+0.0158}_{-0.0048}$ \\
\hline           
     
$r $        &0.0089           &$^{+0.0003}_{-0.0005}$\\
           
\hline
\end{tabular}
\caption{
Constraints on the model parameters and the cosmological parameters. We used the data combination \textit{Planck'18~TT+TE+EE+ low P +lensing + BKPlanck15.} All dimensionful quantities are in reduced Planck  units. In this case no extra dark radiation ($\Delta N_{\rm eff}\approx 0$) is taken as an input.}
\label{T2}
\end{table}
}
\begin{table}[hbt!]
\centering
\begin{tabular}{|m{2cm}  | m{3cm}|   m{2.0cm}|}
\hline
Parameters    &  Central value   & $1\sigma$   \\
\hline
    $R(/10^{-5})$         &     $< 3.05 $     &$--$ \\
\hline

 $V_0(/10^{-11})$         &     7.03  &$^{+3.98}_{-1.52}$   \\
\hline
$A_s(/10^{-9})$       & 1.065         & $^{+0.840}_{-0.230}$   \\
\hline
 $H_0$          & 69.02             &$^{+1.11}_{-1.12}$\\ 
\hline  
     $n_s$          &  0.9830                      & $^{+0.0160}_{-0.0080}$ \\
\hline           
     
$r $        &0.0096           &$^{+0.0002}_{-0.0004}$\\
           
\hline
\end{tabular}
\caption{
Constraints on the model parameters and the cosmological parameters.~We used the data combination \textit{Planck'18~TT+TE+EE+ low P +lensing + BKPlanck15.} All dimensionful quantities are in reduced Planck units. In this case, $\Delta N_{\rm eff}= 0.6$.}
\label{T3}
\end{table}

\begin{table}[hbt!]
\centering
\begin{tabular}{|m{2cm}  | m{3cm}|   m{2.0cm}|}
\hline
Parameters    &  Central value   & $1\sigma$   \\
\hline
    $R(/10^{-5})$         & 2.13           &$^{+0.091}_{-0.051}$ \\
\hline

 $V_0(/10^{-11})$         & 5.70          &$^{+4.10}_{-0.95}$  \\
\hline  

   $\Delta N_{\rm eff}$     &  0.0004         &$^{+0.22}_{-0.19}$ \\
\hline
$A_s(/10^{-9})$       & 1.21          & $^{+0.881}_{-0.250}$   \\
\hline
 $H_0$          & 68.20               &$^{+1.90}_{-3.10}$\\
\hline  
       
  $n_s$          &  0.9701                     & $^{+0.0120}_{-0.0141}$ \\
\hline           
     
$r $        &0.0094            &$^{+0.0004}_{-0.0005}$\\
           
\hline
            
$N_{\rm pivot}$             &53.00           &$^{+1.60}_{-0.50}$\\
 \hline 
 $T_{\rm rh}(/10^{-7})$             &2.00            &$^{+0.65}_{-0.15}$ \\
 \hline 
 
\end{tabular}
\caption{
Constraints on the model parameters and the cosmological parameters using {\sc{Cobaya}}. We used the data combination \textit{Planck'18~TT+TE+EE+ low P +lensing + BKPlanck15.} All dimensionful quantities are in reduced Planck  units.}
\label{T4}
\end{table}
\section{Discussion and Conclusions}
\label{scon}

In the present work, we focused on the phenomenology of fibre inflation. We would like to begin this section by noting some issues related to the construction
of the model in string compactifications. First, there is the possibility of the presence of certain $\alpha'$ corrections \cite{w1,w2} in the effective action (which are still not completely
understood) that might contribute to the positive exponential term in \pref{ipot}. One consequence of this might be that the coefficient of the positive exponential
term can be pushed to higher values. In this case, our analysis would have to be redone, taking into account the appropriate range for
$R$. A similar issue is the geometric instability that can arise as a result of  the ultralight field in the model \cite{geos}. At this stage, it is unclear how relevant the instability
is for fibre inflation, but it could have implications on the parameter space of the model. Apart from the above issues, we also note that the model is in tension with some quantum gravity conjectures which recently gained attention: the Swampland conjectures \cite{sw} and the  Trans-Planckian Censorship conjecture
\cite{cc}. However, the methods in the paper are general enough to be easily modified if there is better understanding of the parameter space of the model and its consistency with such conjectures.

  Our results are interesting from the point of view of phenomenology. As reported in Table~\ref{T1}, the central value of the tensor-to-scalar ratio is $r \sim 0.00932$, which is in the observably verifiable range for the next generation of CMB B-mode surveys. More generally,
the results for fibre inflation in the present article and our earlier work \cite{kcmb}  in the context of K\"ahler moduli inflation show that precision cosmology
can be a powerful tool to constrain string compactifications. As emphasized in Sec.~\ref{sm}, we have carried out our analysis using two different MCMC simulators, the results of which are consistent. This validates the robustness of the numerically computed best-fit values of the model parameters and inflationary observables quoted as our main results in Tables~\ref{T1} - ~\ref{T4}.

As future directions, it will be interesting
to look for top-down constructions of string models, with the model parameters in ranges obtained from our analysis. It will also be interesting to compare 
with the preferred ranges from the point of view of particle physics  \cite{pp1}. Another phenomenologically exciting avenue is to carry out a similar analysis for closely related models,
including the $\alpha$-attractor class \cite{iz}. 

\subsubsection*{Note added} 
Recently, Ref.~\cite{V} was submitted to the arXiv, 
which essentially addressed the same problem as this article. The methods employed and the data sets used in the two articles are different,
yet, broadly, the results agree.

\section*{Ackowledgements}

We would like to thank Michele Cicoli, Fernando Quevedo, L. Sriramkumar, Eleonora Di Valentino,
V. Vennin, and Ivonne Zavala  for discussions. A.M. would like to thank the National Taiwan University and IIT Madras  for hospitality during the course of the project. M.R.G. would like to thank the Saha Institute of Nuclear Physics for hospitality during the course of the project. M.R.G. and S.B. would also like to thank the WHEPP-XVI organizers for their kind hospitality and for providing an atmosphere conducive to discussions. The work of M.R.G. is supported by the Department of Science and Technology, Government of India under Grant Agreement No. IF18-PH-228 (INSPIRE Faculty Award). S.B. is supported by the institute postdoctoral fellowship at Physical Research Laboratory, India. We also acknowledge the computation facility, 100TFLOP HPC Cluster, Vikram-100, at Physical Research Laboratory, Ahmedabad, India.

\end{document}